\documentclass[a4paper,11pt]{article}
\pdfoutput=1 

\usepackage{jheppub} 

\usepackage{comment}
\usepackage[T1]{fontenc} 

\usepackage{gensymb}
\usepackage{graphicx}
\usepackage{caption}
\usepackage{subcaption}

\title{Modification of Fox-Wolfram Moments for Hadron Colliders}

\author{L. A.  Spiller}
\affiliation{University of Melbourne,\\Victoria, Australia}

\emailAdd{lspiller@cern.ch}

\abstract{Collisions of composite particles impose an arbitrary boost in the longitudinal direction on a given event. This implies that the centre-of-mass frame at hadron colliders is undetermined for processes with missing energy in the final state. This motivates the modification of the Fox-Wolfram moments such that the moments for a given event are identical when viewed in the lab or centre-of-mass frame of the beam. The resulting moments are invariant under rotations in the plane transverse to the beam and boosts parallel to the beam. These moments are then used to demonstrate improved signal separation in the channel where the Higgs decays to two b-quarks while being produced in association with a vector boson.}

\begin{document}
\maketitle
\flushbottom

\section{Introduction}
\label{overall:intro}
The complexity of final states being studied at hadron colliders motivates the use of topological variables. The topological variables being used at the Large Hadron Collider (LHC) were often developed in the context of electron-positron colliders. This is true of the Fox-Wolfram moments (FWMs). The FWMs have been applied in the context of the ATLAS and CMS experiments \cite{LHC1,LHC2,LHC3} but are not widely used. In contrast, the B-factories, such as Belle and BaBar, use the FWMs extensively to partition phase space \cite{BELLE1,Babar}. In particular, the FWMs are used in the algorithms to suppress the continuum background \cite{BelleBook}. The FWMs partition phase space in a way which is natural for B-factories but not always natural for hadron colliders. The FWMs correspond to a decomposition of the event's phase space into Fourier modes on the surface of a sphere. In the context of hadron colliders, it would be desirable for a harmonic analysis to be invariant under Lorentz boosts parallel to the direction of the beam. Incorporating and evaluating this change will be the subject of this submission.

\section{Overview of Fox-Wolfram Moments}
\label{sec:intro}
The FWMs are defined as \cite{foxwolfram1,foxwolfram2,foxwolfram3}:
\begin{equation}
\label{eq:foxwolfram}
H_{l} = \frac{4 \pi}{2l + 1} \sum\limits_{m=-l}^{l}\left| \sum\limits_{i=1}^{N} Y_{l}^{m}\left( \theta_i, \phi_i \right) \left| \vec{p}_i \right| \right| ^{2},
\end{equation}
where $Y_l^m$ are the of spherical harmonics, $\sum\limits_{i=1}^{N}$ is the sum over all reconstructed objects or particles, $\sqrt{s}$ is the centre-of-mass energy of the collision and $\theta_i$, $\phi_i$ and $\vec{p}_i$ are the i\textsuperscript{th} object's spherical coordinates and momentum in the event's centre-of-mass frame. The moments can be written in terms of the angular distance between each final state object:
\begin{equation}
\label{eq:altfoxwolfram}
H_{l} = \sum\limits_{i,j=1}^{N} \left| \vec{p}_i \right| \left| \vec{p}_j \right| P_l \left( \cos \Omega_{ij} \right),
\end{equation}
using the addition formula for the spherical harmonics:
\begin{equation}
\label{eq:legendreidentity}
P_l( \cos \Omega_{i,j} ) = \frac{4 \pi}{2l+1} \sum\limits_{m=-l}^{m=l} Y_l^m( \phi_i, \theta_i ) Y_l^{*m}( \phi_j, \theta_j ),
\end{equation}
where $P_l$ is the Legendre polynomial of order $l$ and $\Omega_{ij}$ is the angular distance between particle $i$ and $j$. In this form, the invariance under rotation is made manifest as the dependence on the arbitrary axis in equation \ref{eq:foxwolfram} disappears. The moments are typically normalised to the zeroth moment; this corresponds to a uniform rescaling for electron-positron colliders but is more significant in hadron colliders.

Intuitively, the Fox-Wolfram moments describe how compatible the event topology is with each of the spherical harmonics. The utility of the FWMs arises because the moments form an orthogonal basis and are invariant under the set of rotations, $SO(3)$. The rotational symmetry reflects the symmetry of the particle collision in the event's centre-of-mass frame. In effect, this means that any rotation of a given event will not change the moment (up to and excluding detector effects). Similarly, orthogonality removes the redundancy between different moments. When applied correctly, this allows the most important features of an event to be reduced to the lower order moments, with higher order moments describing features of the event dependent on finer resolutions. The FWMs do not contain enough information to reconstruct the energy-density because all information about phases has been removed. They do, however, allow the reconstruction of the energy-density correlation function.

\section{Limitations of the  Fox-Wolfram Moments}
In hadron collider experiments, the Fox-Wolfram moments no longer describe the symmetry of the detector and are no longer orthogonal. The limitation of hadron colliders is that the centre-of-mass frame of the event cannot be accurately reconstructed for many processes; particularly processes with final states containing missing energy or where objects are mis-measured jets. In $e^+e^-$ colliders, the centre-of-mass energy of the collision is known and the event can always be boosted from the lab frame into the centre-of-mass frame. By contrast, hadron colliders can only determine the transverse missing energy, so for final states with large missing energy, the event's four-vector sum cannot be reconstructed. The use of FWMs in this context implicitly assumes that the event is produced at rest in the lab-frame. This motivates the creation of a new set of moments that, unlike the Fox-Wolfram moments, are invariant under longitudinal boosts rather than rotations in the polar angle, $\theta$. Both moments are invariant under rotations in the transverse plane. For brevity, the new moments will be referred to as the Hadron Collider Moments (HCMs).

A further weakness of the FWMs is that the orthogonality condition for the spherical harmonics,
\begin{equation}
\label{eq:orthog}
\int\int_{\Omega} Y_l^m Y_{l'}^{m'*} d \Omega = 4 \pi \delta_{l l'} \delta_{m m'},
\end{equation}
no longer holds because of the incomplete coverage of the detectors. For example, the ATLAS detector's inner tracker has a coverage in rapidity of $\left| \eta \right| \lesssim 2.5$, corresponding to $\theta \approx 40\degree$ missing out of $360 \degree$. Over this reduced integral, equation \ref{eq:orthog} becomes:
\begin{equation}
\int\int_{\Omega '_{\textrm{cut}} \subset \Omega} Y_l^m Y_{l'}^{m'*} d \Omega \neq 4 \pi \delta_{l l'} \delta_{m m'},
\end{equation}
and hence the spherical harmonics no longer form an orthogonal basis. This limitation also applies to experiments where the centre-of-mass frame can be reconstructed. This problem is identical to that faced by the Cosmic Microwave Background (CMB) mapping experiments where the all-sky spectrum is expressed as a Fourier-Legendre series. In CMB experiments, the galactic plane must be excluded (again approximately $40\degree$ of $360\degree$). To remedy this, a simple orthogonalisation procedure can be adopted identical to that used by the CMB experiments \cite{Orthog}. This limitation is not present in the HCMs.

\section{Symmetries of the Lab Frame in Hadron Colliders}
The natural symmetries of an event in the centre-of-mass frame is $SO(3)$, or the two rotations of $\mathbb{R}^3$. By contrast, the natural symmetries of an event in the lab frame of a hadron collider experiment is $SO(2)_T \times SO(1,1)_\beta$, where $SO(2)_T$ refers to rotations in the plane transverse to the beam and $SO(1,1)_\beta$ refers to the Lorentz boosts parallel to the beam. This motivates the use of the standard coordinate system of hadron colliders:
\begin{equation}
\label{eq:hadroncoordinates}
p = m_T \cosh \left( y \right),\;\;\; p_x = p_T \cos \phi,\;\;\; p_y = p_T \sin \phi,\;\;\; p_z = m_T \sinh \left( y \right),
\end{equation}
where the rapidity, $y$, is defined as:
\begin{equation}
\label{eq:rapidity}
y = \frac{1}{2} \log \frac{E+p}{E-p}.
\end{equation}
The rapidity if often approximated by the pseudorapidity:
\begin{equation}
\label{eq:rapidity}
\eta = - \log \left( \tan \frac{\theta}{2} \right),
\end{equation}
which is valid in the limit that the mass of the particles are negligible relative to their energy. At the LHC this is valid for leptons but may not be valid for jets which often have much larger masses. Similarly, the transverse mass, $m_T$, is often approximated to the transverse momentum, $p_T$, in this limit. This coordinate system has the advantage that the symmetries of the system are manifest with the invariance of both $\Delta \eta$ and $\Delta \phi$ under longitudinal boosts and rotations about the transverse plane.

It is also useful to define an invariant distance in this coordinate system:
\begin{equation}
\label{eq:distance}
\Delta R = \sqrt{\Delta \eta^2 + \Delta \phi^2} = R_1^2 + R_2^2 - 2 R_1 R_2 \cos ( \Delta \gamma ),
\end{equation}
where $\Delta \gamma$ is the angle between the two radii in $\eta$-$\phi$ space. This contrasts with \ref{eq:altfoxwolfram} where the distance on a great circle, $\Omega$, is given by:
\begin{equation}
\label{eq:circledistance}
\cos \Omega_{i,j} = \cos ( \phi_i ) \cos ( \phi_j ) + \sin ( \phi_i ) \sin ( \phi_j ) \cos ( \theta_i - \theta_j  ).
\end{equation}

\section{Proposed Moments}
The FWMs are equivalent to a Fourier series on the surface of a sphere, also known as a Legendre series. Equation \ref{eq:distance} describes a cone with radius $\Delta R$. In this case, it is necessary to take a Fourier series on the radial component of a cylinder, also known as a Fourier-Bessel series. Together with $\gamma$, the polar angle of $R$ in $( y, \phi)$ space, this forms a set of cylindrical coordinates with the metric:
\begin{equation}
\label{eq:metric}
ds^2 = m_T^2 d y^2 + p_T^2 d \phi ^2 \approx p_T^2 \left( d R^2 + R^2 d \gamma^2 \right)
\end{equation}

The new set of moments can be calculated for the new symmetry by solving the Laplacian for the metric in equation \ref{eq:metric}. Alternatively, the Fourier expansion can be rewritten in terms of $\gamma$ and $R$ and the moments inferred. This can be seen from the partial wave expansion:
\begin{equation}
    e^{- i\vec{k} \cdot \vec{r}} = e^{-ikrsin(\theta)} = \sum\limits_{m=-\infty}^{\infty} J_m \left( kr \right) e^{-im\theta} i^m,
\end{equation}
where $J_m$ is the $m^{th}$ Bessel function. The partial wave expansion can be used to rewrite the Fourier series as:
\begin{equation}
    \label{eq:newfourier}
    \sum\limits_{n=1}^{\infty} a_n e^{-i\vec{k_n} \cdot \vec{r} } = \sum\limits_{n=1}^{\infty} \sum\limits_{m=-\infty}^{\infty} J_m \left(k_n r \right) e^{-im\theta} i^m
\end{equation}
As with the FWMs, the HCMs describes the Fourier modes of the density correlation function and not the density. Using the equation \ref{eq:newfourier}, a set of modes can be constructed which are analogous to equation \ref{eq:foxwolfram}:
\begin{equation}
    \label{eq:HCM}
    S_n = \sum\limits_{m=-\infty}^{\infty} \left| \sum\limits_{i=1}^{N} p_{T,i} J_m \left( k_l R \right ) e^{-im \gamma } \right|^2.
\end{equation}
The infinite sum in this equation limits the usefulness of this form. The infinite sum can be removed by applying the addition theorem for Bessel functions,
\begin{equation}
\label{eq:besselidentity}
J_0 ( \Delta R_{1,2} ) = \sum\limits_{m = - \infty }^{ \infty } J_m \left( R_1 \right) J_m \left( R_2 \right) e^{ - i m \Delta \gamma_{1,2} },
\end{equation}
where $\Delta R_{1,2}^2 = R_1^2+R_2^2-2R_1 R_2 \cos \gamma_{i,j}$. The resulting equation has the dependence on an arbitrary axis removed and is analogous to equation \ref{eq:altfoxwolfram}:
\begin{equation}
    S_n = \sum\limits_{i,j} p_{T,i} p_{T,j} J_0 \left( k_l \Delta R_{i,j} \right).
\end{equation}
The coefficients of this expansion, $k_l$, are defined by the boundary conditions. Using the Dirichlet boundary condition that the energy-density correlation function becomes zero outside of the detector acceptance region, the coefficients, $k_{l}$, corresponds to the $l^{th}$ zero of $J_0$. For convenience, $u_{0}$ is set to zero. The moments are normalised to the zeroth HCM in the same manner as the FWMs. The dependence on the arbitrary axes disappears and the symmetries become manifest. Using these new moments, an event can be expected to give the same value whether they are measured in the lab or centre-of-mass frame.

Information about the particle identifications can be incorporated into both the FWMs and the HCMs by splitting up the summand as follows:
\begin{equation}
\label{eq:splitsum}
\sum_{i,j=all} = \sum_{i,j=\text{jets}} + \sum_{i,j=\text{leptons}} 2 \times \sum_{i=\text{jets}, j=\text{leptons}}.
\end{equation}
This allows the moments to be divided into several moments consisting of each of these summands:
\begin{equation}
\label{eq:splitmoments}
S_{l} = S_{l, \text{lepton} \times \text{lepton}} + S_{l, \text{jets} \times \text{jets}} + 2 \times S_{l, \text{leptons} \times \text{jets} }.
\end{equation}

Equation \ref{eq:altfoxwolfram} is a Fourier-Bessel series which can also be interpreted as the Hankel transform on a discrete interval. The Hankel transform is linked to the Abel transform by the Projection-Slice theorem~\cite{abeltransform}. In this theorem, the Hankel transform is equivalent to the Abel transform view in Fourier space:
\begin{equation}
    \mathcal{F} \cdot \mathcal{A} = \mathcal{H}.
\end{equation}
The Abel transform of a function, $f(r)$, can be written in the following form:
\begin{equation}
    \mathcal{A} = \int\limits_{-\infty}^{\infty} f\left(\sqrt{x^2 + y^2}\right) dy.
\end{equation}
This allows the moments to also be interpreted as examining, in Fourier space, the projection of the density correlation function in $(\eta, \phi)$ space along parallel lines of constant $\phi$.

\section{Application to Associated Production of $\text{H} \rightarrow \text{b}\bar{\text{b}}$}
As a simple test of the utility of the HCMs relative to the FWMs, the moments are applied to the all-hadronic final state of the $\text{H} \rightarrow \text{b} \bar{\text{b}}$ produced by Higgs-strahlung from a $W$- or $Z$-boson. This process was chosen because of the complexity of its final state, and it is a new channel which offers the potential to increase the explored final states of the process $\text{H} \rightarrow \text{b}\bar{\text{b}}$ at the LHC. One of the main backgrounds for this signal is the production of $t\bar{t}$ and $t\bar{t} + \text{jets}$. The moments are tested for their discriminating power against this background.

\begin{figure}[Ht]
\centering
\includegraphics[width=.9\textwidth]{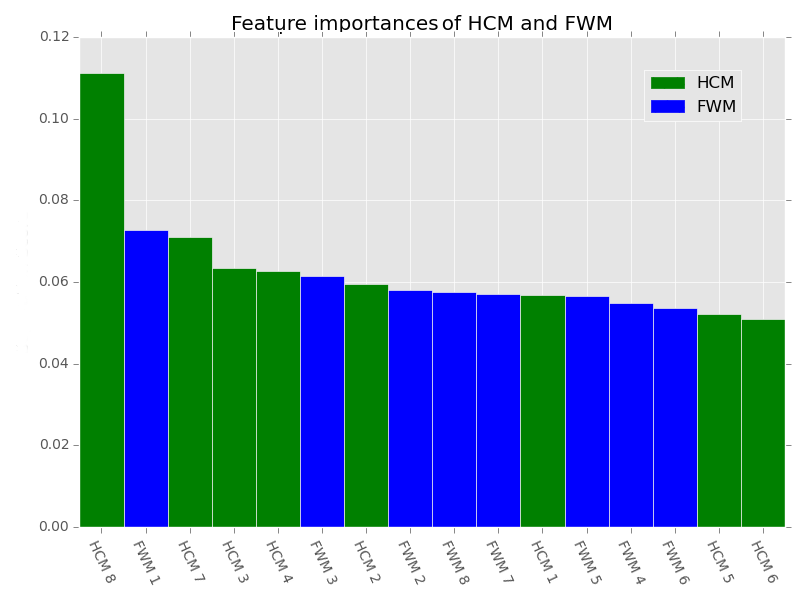}
\caption{\label{fig:i} The distribution of feature improtance from the random forest when testing the separation of the all-hadronic final state for the signal $\text{H} \rightarrow \text{b}\bar{\text{b}}$, produced in association with a $W$- or $Z$-boson, from the backgrounds consisting of top quark pair production with additional jets. The distribution of feature importances is shown for the FWMs (blue) and the HCMs (green) with the moments arranged in decreasing importance. The feature importance in each tree is defined as the normalised reduction in node impurity brought by all splits on that feature in the tree. The overall features importance is the average feature importance across all trees. For brevity, only the first eight moments were calculated for both the FWMs and HCMs. The corresponding index for each moment is shown on the x-axis.}
\end{figure}

\begin{figure}[Ht]
\centering
\includegraphics[width=.85\textwidth]{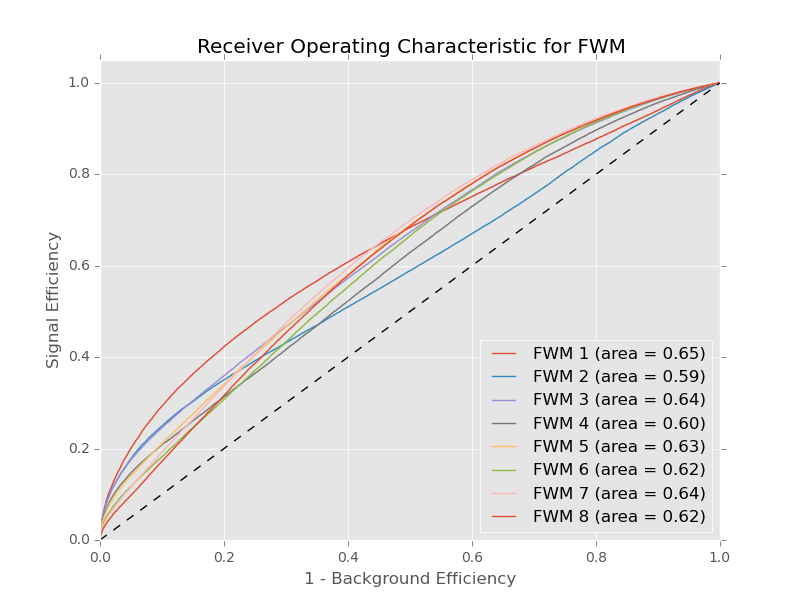}
\hfill
\includegraphics[width=.85\textwidth]{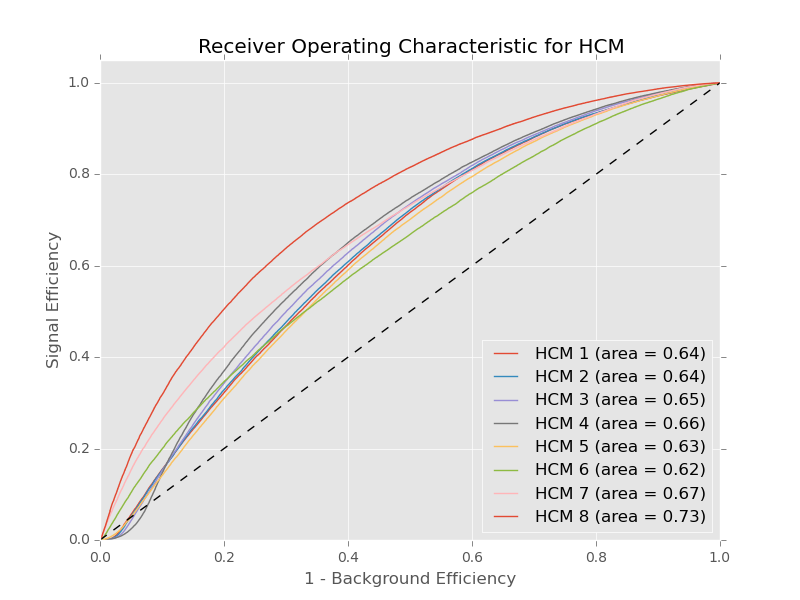}
\caption{\label{fig:j} The ROC curves and their respective areas for the FWMs (top) and HCMs (bottom) used in testing the separation of the all-hadronic final state for the signal $\text{H} \rightarrow \text{b}\bar{\text{b}}$, produced in association with a $W$- or $Z$-boson, from the backgrounds consisting of top quark pair production with additional jets.}
\end{figure}

\begin{figure}[Ht]
\centering 
\includegraphics[width=.6\textwidth]{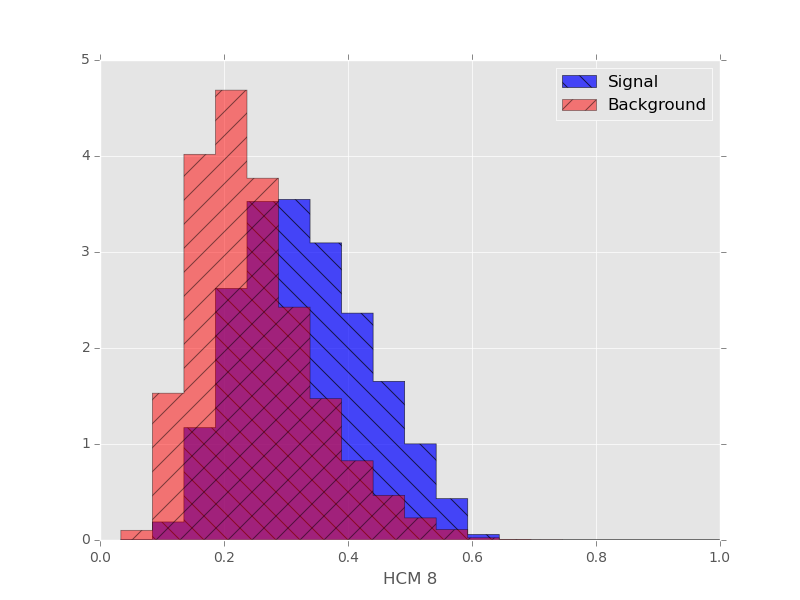}
\caption{\label{fig:k} The distribution for the eighth order HCM ($l=8$) for the all-hadronic final state for the signal $\text{H} \rightarrow \text{b}\bar{\text{b}}$ (blue), produced in association with a $W$- or $Z$-boson, from the backgrounds consisting of top quark pair production with additional jets (red). This moment was shown to have the most separation of the HCMs (for $l<8$).}
\end{figure}
\subsection{Simulation and Monte-Carlo Generation}
The simulated events used to test the new moments are generated using the \texttt{MadGraph} \cite{MADGRAPH} and \texttt{Pythia8} \cite{PYTHIA} simulation framework with the ATLAS detector simulated by Delphes \cite{Delphes} with proton-proton collisions at $\sqrt{s}=13$ TeV. A filter is placed on the jets which corresponds to a 17 GeV cut on jet $p_T$ at the parton level and a 20 GeV cut was used at reconstruction level. The background sample includes up to three additional jets. The final state requires at least two jets which are tagged as coming from the decay of a b-quark.

\subsection{Quantifying Separation}
The separation of these two processes is quantified by the integrated area under the receiver operating characteristic (ROC) curves. In addition to the ROC curves, a simple random forest~\cite{SCIKIT} is constructed to obtain the separation which includes the effect of correlations. A random forest was chosen for this task because it is able to more fully explore the full phase space provided by these variables in a less biased way. At each decision node in the decision tree, the best cut from a random subset of the input variables is performed; by contrast a classic decision tree is usually deterministic. Even when boosted, a decision tree will tend to look fairly uniform and it will consistently cut on a few dominant moments in the first few decision nodes. This prevents the full phase space of the moments being fully explored.

The feature importance is generated by the random forests and consists of the normalised reduction in node purity brought by all splits on that feature averaged across all trees in the random forest (with the sum of all scores normalised to one)~\cite{SCIKIT}. The distribution of feature importances for the first eight moments of the FWMs and HCMs are shown in figure \ref{fig:i}.

\subsection{Analysis}
Figure \ref{fig:i} shows greater separation between signal and background when using the HCMs over the FWMs for the first eight moments. The moment which demonstrates the most separation for the HCMs are shown in Figure \ref{fig:k} and the feature importance and ROC curves for the HCMs and FWMs are shown in \ref{fig:i} and \ref{fig:j} respectively. The HCMs, in particular the eighth moment, show consistently better performance and separation than the FWMs. This is reflected by both the random forest's ranking of feature importance and the ROC curves.

 The use of the FWMs and the HCMs are not mutually exclusive. Both provide information based on two different limits. The FWMs approximate the final state as being produced at rest while the HCMs assume that the boost is arbitrary. For the very heavy final mass states, the FWMs are expected to improve because heavier mass states are generally very close to rest in the rest frame of the two colliding protons.

\section{Conclusion}
The inability to determine the centre-of-mass frame of the collision in events with missing energy causes the centre-of-mass to only be determinable in the transverse plane. This causes the standard rotational symmetries implicit in the Fox-Wolfram moments to no longer hold. Hence, the same underlying event topology can lead to different Fox-Wolfram moments due to the varying longitudinal boost. To remedy this, a new set of topological moments are proposed for hadron colliders which reflect the underlying symmetry of the lab frame. The use of HCMs and FWMs is complementary with both moments representing limiting cases. The HCMs can be incorporated into analyses in the same way as the FWMs are; individual moments can be used as cuts or inputs for a multi-variable analysis.



\acknowledgments

The authour gratefully acknowledges Noel Dawe, Elisabetta Barberio and Daniele Zanzi, for their help writing and reviewing this submission.

\end{document}